\documentclass{jfm}

\usepackage{graphicx,amsmath,amssymb,color,natbib}
\ifCUPmtlplainloaded \else
  \checkfont{eurm10}
  \iffontfound
    \IfFileExists{upmath.sty}
      {\typeout{^^JFound AMS Euler Roman fonts on the system,
                   using the 'upmath' package.^^J}%
       \usepackage{upmath}}
      {\typeout{^^JFound AMS Euler Roman fonts on the system, but you
                   dont seem to have the}%
       \typeout{'upmath' package installed. JFM.cls can take advantage
                 of these fonts,^^Jif you use 'upmath' package.^^J}%
      }
  \else
  \fi
\fi

\ifCUPmtlplainloaded \else
  \checkfont{msam10}
  \iffontfound
    \IfFileExists{amssymb.sty}
      {\typeout{^^JFound AMS Symbol fonts on the system, using the
                'amssymb' package.^^J}%
       \usepackage{amssymb}%
       \let\le=\leqslant  
         
      }{}
  \fi
\fi

\providecommand\bphi{\boldsymbol{\phi}}
\providecommand\bpsi{\boldsymbol{\psi}}
\providecommand\bq{\mathbf{q}}
\providecommand\bu{\mathbf{u}}

\providecommand\bx{\mathbf{x}}

\providecommand\tphi{\skew4\tilde{\boldsymbol{\phi}}}
\providecommand\cphi{\skew4\check{\boldsymbol{\phi}}}

\providecommand\eps\epsilon
\providecommand\del\delta
\providecommand\p\partial

\renewcommand\i{\text{i}}

\ifCUPmtlplainloaded \else
  \IfFileExists{amsbsy.sty}
    {\typeout{^^JFound the 'amsbsy' package on the system, using it.^^J}%
     \usepackage{amsbsy}}
    {\providecommand\boldsymbol[1]{\mbox{\boldmath $##1$}}}
\fi

\begin{document}

\title{Identifying the active flow regions that drive linear and nonlinear instabilities}
\author{Olivier Marquet$^1$ \& Lutz Lesshafft$^2$}
\affiliation{$^1$ONERA/DAFE, 8 rue des Vertugadins, 92190 Meudon, France\\
$^2$Laboratoire d'Hydrodynamique, CNRS/\'Ecole polytechnique,
91128 Palaiseau, France}
\date{\today}
\maketitle
\author{Olivier Marquet \& Lutz Lesshafft}

\abstract{
A new framework for the analysis of unstable oscillator flows is explored. In linear settings, temporally growing perturbations in a non-parallel flow represent unstable eigenmodes of the linear flow operator. In nonlinear settings, self-sustained periodic oscillations of finite amplitude are commonly described as nonlinear global modes. In both cases the flow dynamics may be qualified as being \emph{endogenous}, as opposed to the exogenous behaviour of amplifier flows driven by external forcing. This paper introduces the \emph{endogeneity} concept, a specific definition of the sensitivity of the global frequency and growth rate with respect to variations of the flow operator. The endogeneity, defined both in linear and nonlinear settings, characterizes the contribution of localized flow regions to the global eigendynamics. It is calculated in a simple manner as the local point-wise inner product between the time derivative of the direct flow state and an adjoint mode. This study demonstrates for two canonical examples, the Ginzburg--Landau equation and the wake of a circular cylinder, how an analysis based on the endogeneity may be used for a physical discussion of the mechanisms that drive a global instability. The results are shown to be consistent with earlier `wavemaker' definitions found in the literature, but the present formalism enables a more detailed discussion: a clear distinction is made between oscillation frequency and growth rate, and individual contributions from the various terms of the flow operator can be isolated and separately discussed. In particular, in the context of nonlinear saturated oscillations in the cylinder wake, such an analysis allows to discriminate between the quasi-linear dynamics of fluctuations around a time-averaged mean flow on one hand and the effect of harmonic interactions on the other hand; the results elucidate why a linear analysis of the mean flow in this particular case provides accurate predictions of the nonlinear dynamics.
}

\section{Introduction}

Global instability in flows denotes the possibility of a spontaneous bifurcation from a steady flow state to a time-periodic state of synchronised oscillations in the entire flow field.  A commonly observed scenario is that of a supercritical Hopf bifurcation, where linearly unstable perturbations of small amplitude first experience exponential growth, until nonlinear effects lead to  amplitude saturation. The final time-periodic flow state is named a nonlinear global mode \citep{HM90}; the exponentially growing small-amplitude perturbations in the early stage of the bifurcation correspond to eigenmodes of the linearized flow operator, traditionally called linear global modes. The attribute \emph{global} is used here to designate an analysis that resolves all non-homogeneous flow directions, as opposed to a \emph{local} Ansatz, which implies the approximation of locally parallel flow. 

Nonlinear global modes are usually obtained as asymptotic oscillatory states from time-resolved numerical simulations, whereas linear global mode analysis requires the solution of linear eigenvalue problems. Complex eigenvalues represent the temporal growth rate and the oscillation frequency; the associated eigenfunctions characterize the spatial distribution of fluctuation amplitude and phase. Linear global mode analysis is now routinely applied to two- and three-dimensional flow configurations. Beyond the primary question whether or not perturbations at small amplitude are unstable, a physical discussion of linear global modes is usually centered around two questions: (i) what are the physical mechanisms that give rise to unstable growth, and (ii) by what means can instability be reduced or enhanced? The first of these questions addresses the endogenous (or intrinsic) flow behaviour, the second question concerns the control of those dynamics through exogenous (or extrinsic) manipulation. 

\cite{HM90} describe the conceptual notion of a `wavemaker' \citep[a word first used by][]{M90} as the region where instability waves are intrinsically generated in globally unstable flows. The interpretation by \cite{K85} of global instability in a wake already uses the same principal idea. \cite{CHR91} derive a formal criterion for the global frequency selection in the context of the linear Ginzburg--Landau equation,  based on the local absolute instability properties. Their formalism is rooted in a WKBJ approximation of instability wavepackets developing in a weakly non-parallel open flow. Within this approximation, local instability waves with upstream- and downstream-oriented group velocity emanate from a streamwise station, the `wavemaker' location, where the two mode branches can be matched by means of a non-physical analytic continuation of the dispersion relation, defined as a function of a complex spatial $x$-coordinate. Such intrinsically generated waves grow and decay as they propagate. While the localized `wavemaker' selects the frequency and drives the global instability mode, it is in general not characterized by large oscillation amplitudes. The spatial separation of the region where waves are generated and the region where they reach their maximal amplitude is caused by convective instability mechanisms in a local sense, or by the non-normality of the linear Navier--Stokes operator in a global sense \citep{CC97}.

A quantitative theory of frequency selection in \emph{nonlinear} systems, still based on the assumption of slow streamwise flow development and for the Ginzburg--Landau model equation, has been proposed by \cite{CC97a} and by \cite{PHCC98}. These studies draw on the theory of front dynamics \citep{S88,S89}, leading to the simple criterion that the nonlinear global mode frequency is given to first order by the absolute frequency at the upstream boundary of a (locally) absolutely unstable flow region of finite extent.  Subsequent applications to wake flows \citep{PH01,P02, C03} suggest that the accuracy of this criterion is only limited by the non-parallelism of the base flow over the distance of amplitude saturation. The transition from (upstream) convective to (downstream) absolute instability marks the `wavemaker' location within the framework of this nonlinear model. 

Linear global modes in non-parallel flows may now be computed directly, without the need for the hypothesis of slow streamwise development. However, the notion of a cause-and-effect relation between different streamwise regions is lost along with this approximation, and the localisation of a `wavemaker' region within a global structure must be accomplished through new criteria. The \emph{sensitivity} of the linear eigenvalue (frequency and growth rate) with respect to localized changes of the flow operator provides the appropriate concepts for a formal definition of a global `wavemaker'. Yet the sensitivity problem may be posed in several ways, depending on the physical premise of what `drives' an instability.

\cite{GL07} provide a discussion of the cylinder wake instability based on the structural sensitivity of the unstable linear eigenmode.  The structural sensitivity, in their definition, quantifies how an eigenvalue is affected by the introduction of localized forcing of a given perturbation quantity, proportional in strength to the same or another perturbation quantity. It thereby provides a measure in every point in space for the effect of internal feedback between perturbations.  \cite{GL07} conjecture that those regions where an altered coupling induces the strongest change of the eigenvalue must also be the most significant regions for the action of internal feedback mechanisms that underpin the genuine eigenmode dynamics. At present, this formalism is arguably the most commonly accepted definition of the `wavemaker' in a global analysis framework. The concept is quite naturally extended to nonlinear global modes by way of Floquet theory \citep{LGP08}. However, one inconvenience of this approach is that it does not distinguish between frequency and growth rate, as the Cauchy--Schwarz theorem is invoked in order to define an upper bound for the drift of the \emph{modulus} of the eigenvalue. Another stems from the large number of feedback relations between the various flow quantities that may prove to be significant. The formalism does not allow to single out the influence of specific terms in the flow equations.

\cite{MSJ08} investigate the sensitivity of the linear cylinder wake instability with respect to localized modifications of the base flow. As far as linear instability is linked to the interaction between perturbations and the base flow, it may be argued that such an analysis is well suited to identify the principal flow regions where instability originates. The formalism distinguishes between frequency and growth rate, yet it is clearly cast in the form of a \emph{control} problem. The question how an instability mode is affected by \emph{exogenous} alterations, be it alterations of the internal feedback \citep{GL07} or of the base flow \citep{MSJ08}, is conceptually different from the question how its \emph{endogenous} dynamics come into being.

The objective of the present paper is to propose a variant of the sensitivity problem for linear as well as nonlinear global modes that identifies more directly those endogenous eigendynamics. For the sake of comparison and validation, these concepts are demonstrated for the two traditional test settings used in  the literature on wavemakers: the one-dimensional Ginzburg--Landau equation and the two-dimensional cylinder wake. It is hoped that the formalism will be useful for the analysis of physical instability phenomena in a wide range of applications. 

The paper is organised as follows. Section 2 documents linear and nonlinear global mode results for the Ginzburg--Landau equation and for the cylinder wake.  This section does not contain genuinely new results, but rather serves as a repertory and review. The configurations discussed here are used in the following as examples in order to demonstrate the proposed formalism. The endogeneity concept is introduced for linear settings in \S{}3, and its application for the analysis of linear global modes is demonstrated for the two example configurations. The extension of the formalism to fully nonlinear situations is laid out in \S{}4. Conclusions are given in \S{}5. 
An appendix addresses the implications of general inner products for the analysis. 

\section{Linear, nonlinear, direct  and adjoint global modes}
\label{sec:global_modes}

The evolution  equation   of a flow variable $\bq(\bx,t)$ is considered in the general form
\begin{equation}
\mathcal{B} \, \partial_t \bq = \mathcal{N}(\bq) \, ,
\label{eqn:nonlinear}
\end{equation}
where $\mathcal{N}$ is a nonlinear operator, and $\mathcal{B}$ is an operator of very simple structure  that indicates  on which component  
of $\bq$ the time derivative applies. In what follows, it will always be assumed that $\mathcal{B}$ is self-adjoint.\\

A base flow $\bq_b$ is  a steady  solution of the nonlinear equation, $\mathcal{N}(\bq_b)=0$. The linear stability of such a steady base flow is investigated by superposing small-amplitude time-dependent perturbations $\bq'$. The dynamics of these perturbations is governed by the linear equation
\begin{equation}
\mathcal{B} \, \partial_t \bq' = \mathcal{L}_{\bq_{b}} \bq'\, ,
\label{eqn:linear}
\end{equation}
where the linear operator $\mathcal{L}_{\bq_{b}}$ is obtained  as the linearisation of $\mathcal{N}$ around the base flow: $\mathcal{N}(\bq_b+\epsilon\bq')=\mathcal{N}(\bq_b) + \epsilon \mathcal{L}_{\bq_b}\bq'$. Eigenvalues $\lambda_j$ and associated eigenfunctions $\bphi_j(\bx)$ of $\mathcal{L}_{\bq_b}$ are obtained as solutions of the eigenvalue problem   
\begin{equation}
\lambda_j  \, \mathcal{B} \, \bphi_{j} = \mathcal{L}_{\bq_{b}} \bphi_j \, .
\label{eqn:eig}
\end{equation}
The eigenmodes  $\bq_j'(\bx,t) = \bphi_j(\bx) \exp( \lambda_j t)$ form  a complete basis for the range of $\mathcal{L}_{\bq_b}$.
In physical terms, the real and imaginary parts of a complex eigenvalue represent temporal growth rate $\sigma$ and frequency $\omega$ of a linear eigenmode. The convention $\lambda=\sigma-\i\omega$ is adopted here, i.e.~the frequency is given by the negative imaginary part of $\lambda$. 

For the purpose of deriving an adjoint equation associated with the direct equation (\ref{eqn:linear}), the following inner product of vector-valued  functions is introduced:
\begin{equation}
\left\{ \boldsymbol{f}(\bx,t), \boldsymbol{g}(\bx,t) \right\} = \int_0^T \!\!\! \int_\Omega \boldsymbol{f}^*(\bx,t) \cdot \boldsymbol{g}(\bx,t) \, \text{d}\bx \, \text{d}t \, ,
\label{eqn:spatiotemp}
\end{equation}
where the star denotes the complex conjugate. The adjoint linear operator $\mathcal{L}^{\dagger}_{\bq_b}$ is  then  derived by requiring
\begin{equation}
\left\{ \bq^\dagger , \mathcal{B} \, \partial_t\bq' - \mathcal{L}_{\bq_{b}} \bq' \right\} = \left\{ -\mathcal{B} \, \partial_t\bq^\dagger - \mathcal{L}^\dagger_{\bq_{b}} \bq^\dagger , \bq' \right\}  .
\end{equation}
The adjoint equation  associated with  (\ref{eqn:linear}) is thus  found as   
\begin{equation}
-\mathcal{B} \, \partial_t \bq^\dagger = \mathcal{L}^\dagger_{\bq_b} \bq^\dagger \, .
\label{eqn:linear_adjoint}
\end{equation}
The linear adjoint operator has eigenmodes $\bq_j^{\dagger}(\bx,t) = \bphi_j^{\dagger}(\bx) \exp(\lambda_j^* t)$. Adjoint eigenvalues $\lambda^*_j$ are the complex conjugate  values  of the corresponding direct eigenvalues $\lambda_j$.
Adjoint eigenfunctions  $\bphi_j^{\dagger}(\bx)$  satisfy  
\begin{equation}
\lambda_j^{*}  \, \mathcal{B} \, \bphi^{\dagger}_{j} = \mathcal{L}^{\dagger}_{\bq_b} \bphi_j^{\dagger} \, .
\label{eqn:eigadj}
\end{equation}
The direct and adjoint eigenfunctions form biorthogonal sets, 
\begin{equation}
\langle \bphi_j^\dagger , \mathcal{B} \bphi_k \rangle = \delta_{jk} \, ,
\label{eqn:biorthogonal}
\end{equation}
 with respect to the purely spatial inner product
\begin{equation}
\langle \boldsymbol{f}(\bx), \boldsymbol{g}(\bx) \rangle = \int_\Omega \boldsymbol{f}^*(\bx) \cdot \boldsymbol{g}(\bx) \, \text{d}\bx \, .
\label{eqn:spatialprod}
\end{equation}

In the definition of \cite{HM90}, a \emph{nonlinear} global mode is a time-periodic solution of the nonlinear evolution equation
(\ref{eqn:nonlinear}),  denoted here by $\bq_0(\boldsymbol{x},t)$. In the examples considered in the present investigation, such a nonlinear global mode is the limit-cycle solution that is reached as the result of amplitude saturation of an initially growing linear eigenmode. The nonlinear global mode has a real-valued global frequency $\omega_g = 2\pi/T$, with $T$ being the cycle period. 

The linear stability of this time-periodic solution is investigated by considering the temporal evolution of small perturbations $\bq'(\bx,t)$, governed by the linear equation
\begin{equation}
\mathcal{B} \, \partial_t \bq'(\boldsymbol{x},t) = \mathcal{L}_{\bq_0(t)} \; \bq'(\boldsymbol{x},t) \, .
\label{eqn:tangential}
\end{equation}
The tangential linear operator $\mathcal{L}_{\bq_0(t)}$ is obtained by  linearizing  the nonlinear operator $\mathcal{N}$ around the time-periodic solution $\bq_0(t)$. As $\mathcal{L}_{\bq_0(t)}$ is also time-periodic, a fundamental set of solutions to (\ref{eqn:tangential}) is given by $\bpsi_j(\bx,t) \exp(\zeta_j t)$ \citep[see][]{IJ90}. The  Floquet modes  $\bpsi_j(\bx,t)$ are $T$-periodic in time, and the associated Floquet multipliers $\exp (\zeta_j T)$ characterize the temporal growth or decay of a mode over one cycle period. The real part of $\zeta_j$ is in fact the Lyapunov exponent, the imaginary part of $\zeta_j$ corresponds to a variation of the fundamental frequency.

An interesting property is that 
 the time derivative of the time-periodic solution,  $\partial_{t} \bq_0$,  is a neutral Floquet mode of the autonomous periodic operator $\mathcal{L}_{\bq_0(t)}$.  This  property results from the phase invariance of the time-periodic solution. 
 Both  $\bq_0(\bx,t)$ and $\bq_0(\bx,t+\delta t)$ represent $T$-periodic solutions of the nonlinear equation (\ref{eqn:nonlinear}), therefore their difference $\delta \bq = \bq_0(\bx,t+\delta t)-\bq_0(\bx,t)$ is $T$-periodic as well. A Taylor expansion for small $\delta t$ gives $\delta \bq= \partial_t \bq_0(\bx,t) \delta t$. It follows that $\partial_t \bq_0(\bx,t)$ is a $T$-periodic solution of \eqref{eqn:tangential}, and is therefore a \emph{neutral} Floquet mode:
\begin{equation}
\bpsi_1 (\bx,t) = \partial_t \bq_0 (\bx,t) \, , \quad \zeta_1 = 0 \, .
\label{eqn:floquetmode}
\end{equation}
An adjoint tangential operator $\mathcal{L}_{\bq_0(t)}^\dagger$ can also be  defined  by requiring
\begin{equation}
\big\{-\mathcal{B} \, \partial_t \bq^\dag - \mathcal{L}^\dagger_{\bq_0(t)}\bq^\dagger , \bq' \big\}= 
\big\{ \bq^{\dagger} , \mathcal{B} \, \partial_t \bq' - \mathcal{L}_{\bq_0(t)} \bq' \big\} \, ,
\end{equation}
 from where follows the adjoint equation associated with (\ref{eqn:tangential}),
\begin{equation}
\label{eqn:floquet_adjoint}
-\mathcal{B} \, \partial_t \bq^\dagger (\bx,t)= \mathcal{L}_{\bq_0(t)}^\dagger \bq^\dagger (\bx,t) \, .
\end{equation}
 The linear operator  $\mathcal{L}^\dagger_{\bq_0(t)}$ is also $T$-periodic. Equation (\ref{eqn:floquet_adjoint})  has fundamental solutions in the form of adjoint Floquet modes $\bpsi^\dagger_j(\bx,t) \exp(\zeta^*_j t)$. The sets $\bpsi^\dagger_j$ and $\bpsi_k$ are again biorthogonal,  with respect to the spatio-temporal inner product (\ref{eqn:spatiotemp}), such that with a suitable normalisation they fulfill 
\begin{equation}
\{ \bpsi^\dagger_j, \mathcal{B}  \bpsi_k \} = \delta_{jk} \, .
\end{equation}
 Only  the adjoint Floquet mode $\bpsi^\dag_1$ associated with the neutral Floquet mode $\bpsi_1$ and $\zeta_1=0$ will be used in the following analysis.

\subsection{Ginzburg--Landau equation}
\label{sec:GL}

The Ginzburg--Landau equation has often served as a simple model for flow instability dynamics \citep{H00}. Its scalar state variable $q(x,t)$ only depends on one spatial coordinate, yet its dispersion relation permits a double branch point that allows to distinguish between absolutely and convectively unstable situations, analogous to open shear flows. The nonlinear Ginzburg--Landau equation is used here in the form
\begin{equation}
\partial_t q = \mathcal{N}(q) =  - U\partial_x q + \mu(x) q + \gamma\partial_{xx} q - \beta |q|^2 q  \, .
\label{eqn:GL_nonlin}
\end{equation}
The operator $\mathcal{B}$, in the general notation (\ref{eqn:nonlinear}), in this case is simply the identity, and the operator $\mathcal{N}$ is given by the right-hand side in (\ref{eqn:GL_nonlin}). It is composed of terms representing convection, linear reactive sources, diffusion, and cubic nonlinearity.
Following \cite{CC97}, constant convection and diffusion parameters are chosen, $U=6$ and $\gamma=1-i$, while a parabolic variation of the reactive parameter is prescribed as
\begin{equation}
\mu(x) = \mu_0 + 0.5 \mu_2 x^2 \, .
\end{equation}
Different values of $\mu_0$ will be used in the following, while $\mu_2=-0.1$ is maintained throughout. This variation yields a strong local stability of the system far from $x=0$. All following calculations are performed on an interval $x\in[-40,40]$, and in all cases the fluctuation amplitudes are indeed negligibly small at the numerical boundaries. 
The nonlinearity parameter is chosen as $\beta=1-\i$, except when linear situations $\beta=0$ are considered.

\subsubsection{Linear Ginzburg--Landau}
\label{sec:linGL}

Note that the zero state $q\equiv 0$ is a steady solution of the nonlinear Ginzburg--Landau equation (\ref{eqn:GL_nonlin}). Linearisation around that state yields the linear Ginzburg--Landau equation 
\begin{equation}
\partial_t q = \mathcal{L} q =  - U\partial_x q + \mu(x) q + \gamma\partial_{xx} q  \, ,
\label{eqn:GL_lin}
\end{equation}
which is equivalent to (\ref{eqn:GL_nonlin}) with $\beta=0$. A subscript 0 could be attached to $\mathcal{L}$, denoting the zero base state, but this will be omitted in the following. The associated adjoint operator is obtained as
\begin{equation}
\mathcal{L}^\dag q^\dag = U\partial_x q^\dag + \mu(x) q^\dag + \gamma^*\partial_{xx} q^\dag \, .
\end{equation}

\begin{figure}
\centering
\includegraphics[width=0.7\textwidth]{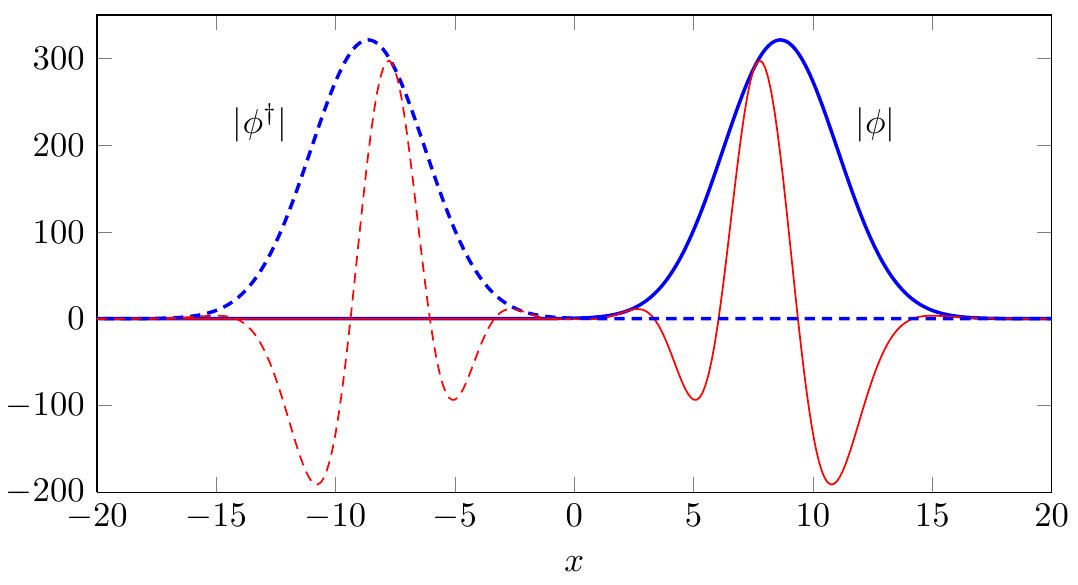}
\caption{Direct and adjoint eigenfunctions (\ref{eqn:GL_lin_direct}, \ref{eqn:GL_lin_adjoint}) of the linear Ginzburg--Landau equation. Absolute values are shown as thick blue lines, real parts as thin red lines. Solid lines represent the direct mode, dashed lines represent the adjoint mode. The scaling is such that $\langle \phi^\dagger,\phi\rangle = 1$.}
\label{fig:eigenfunctions_linGL}
\end{figure}

As derived by \cite{CHR87}, the leading eigenmodes of the direct problem $\lambda \phi = \mathcal{L}\phi$ and of the adjoint  problem  $\lambda^* \phi^\dag = \mathcal{L}^\dag\phi^\dag$ are found as
\begin{align}
\lambda &= \mu_0 - \frac{U^2}{4\gamma} - \sqrt{\frac{-\mu_2\gamma}{2}} \, , \\
\phi(x) &= \exp\left( \frac{Ux}{2\gamma} - \sqrt{\frac{-\mu_2}{2\gamma}}\frac{x^2}{2} \right)\, , \label{eqn:GL_lin_direct} \\
\phi^\dagger(x) &= \exp\left(- \frac{Ux}{2\gamma^*} - \sqrt{\frac{-\mu_2}{2\gamma^*}}\frac{x^2}{2} \right)\, .
\label{eqn:GL_lin_adjoint}
\end{align}
For fixed values of $U$, $\gamma$ and $\mu_2$,
the real-valued parameter $\mu_0$ completely determines the eigenvalue, but it has no influence on the eigenfunction shape. The direct and adjoint eigenfunctions are shown in figure \ref{fig:eigenfunctions_linGL}.  Their maxima are located at $x=8.6$ and  $x=-8.6$, respectively. According to \cite{CC97}, global instability arises when $\mu_0>\mu_c$, with a critical value
\begin{equation}
\label{eqn:mu_c}
\mu_c = \frac{U^2}{4|\gamma|^2} + \bigg|\sqrt{\frac{-\mu_2 \gamma}{2}}\bigg| \cos \bigg(\frac{\arg\,\gamma}{2}\bigg) \, .
\end{equation}
The local instability properties are given by \cite{CHR88}. The model is locally stable wherever $\mu(x)<0$, convectively unstable for $0 < \mu(x) < U^2/4|\gamma|^2$ and absolutely unstable for $\mu(x) > U^2/4|\gamma|^2$. The extent of these regions depends on the value of $\mu_0$.

\subsubsection{Nonlinear Ginzburg--Landau}
\label{sec:nonlinGL}

\cite{PHCC98} report that the Hopf bifurcation in a nonlinear Ginzburg--Landau system is supercritical.  Nonlinear global instability therefore follows from linear global instability, $\mu_0>\mu_c$, and the nonlinear global mode emerges after saturation of the growing linear global mode. A strongly supercritical setting is considered here, with  $\mu_0=2\mu_c$.

The direct nonlinear global mode is computed by time-stepping (\ref{eqn:GL_nonlin}), starting from the linear global mode shape (\ref{eqn:GL_lin_direct})
 as an initial perturbation at low amplitude, until an asymptotic time-periodic state $q_0$  is reached. The linearisation of Equation (\ref{eqn:GL_nonlin}) around the oscillating state $q_0$ is accomplished as detailed in \cite{H15}. The nonlinear term $\beta |q|^2 q $ is properly linearized by augmenting the complex state variable with its complex conjugate, i.e. writing out the nonlinear and linear equations in terms of state vectors $(q, q^*)^T$. In the present study, the direct linear equation is in fact never needed, except for the derivation of the adjoint linear equation, 
\begin{equation}
-\partial_t q^\dagger -U\partial_x q^\dagger - \mu q^\dagger - \gamma^*\partial_{xx} q^\dagger - 2 \beta^* |q_0(t)|^2 q^\dagger + \beta^* q_0^2(t) q^{\dagger *} = 0 \, .
\label{eqn:GL_adjoint}
\end{equation}
With the known time-periodic solution $q_0(t)$, this equation is numerically integrated  backward in time, until convergence towards a periodic solution $q^\dagger_1$ is reached. The existence of such an asymptotic solution is guaranteed, because this is the adjoint Floquet mode associated with the neutral direct Floquet mode $q_1$, defined as the time-derivative of the nonlinear time periodic solution $q_1 = \partial_t q_0$. Details on Floquet theory can be found in \cite{IJ90}.

A Crank--Nicolson scheme is used for the time integration of the nonlinear direct and the associated linear adjoint problem. The use of upwinding finite-difference stencils (seven points) for the spatial discretisation of the direct problem, and downwinding for the adjoint problem, is essential in order to achieve the required numerical accuracy.

\begin{figure}
\centering
\includegraphics[width=\textwidth]{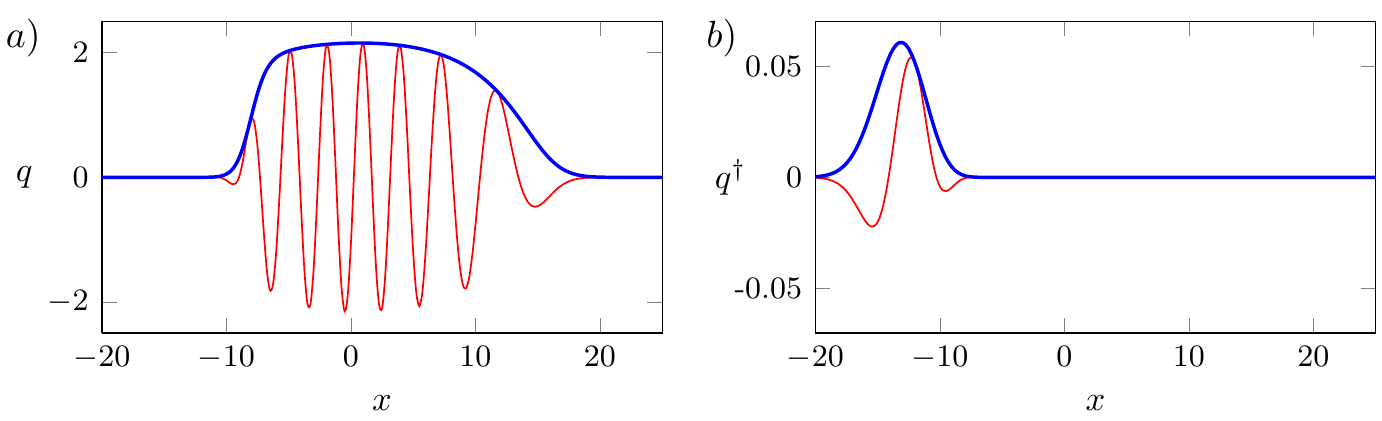}
\caption{{\em a}) Direct nonlinear global mode of the Ginzburg--Landau equation, with parameters as given in the text. Thick blue line: amplitude envelope; thin red line: snapshot of the real part of $q$. {\em b}) Associated neutral adjoint Floquet mode. Same line styles as in {\em a}.}
\label{fig:eigenfunctions_nonlinGL}
\end{figure}

Figure \ref{fig:eigenfunctions_nonlinGL}({\em a}) shows the nonlinear global mode $q_0$ as it is recovered after transients have disappeared. The amplitude envelope has the emblematic shape of an `elephant' mode \citep{PHCC98}, with a sharp upstream wavefront and a softer downstream decay. According to WKBJ theory, the front should be situated at the upstream boundary of the absolutely unstable region in $x$. In the present case, absolute instability prevails in the interval $x \in [-10,10]$, and the foot of the front is indeed placed around $x=-10$. The adjoint Floquet mode, represented in figure \ref{fig:eigenfunctions_nonlinGL}({\em b}), has significant amplitudes only upstream of the direct wavepacket, with a maximum near $x=-13$.

It is to be noted that the nonlinear global mode of the Ginzburg--Landau equation only contains one single frequency, because the nonlinear term is of such a form that it does not generate harmonics. As a result, the time signal in each point shows pure sinusoidal oscillations with zero mean. This property proves to be very convenient for all further discussion, whereas the harmonics and the non-zero mean oscillations that are characteristic for the nonlinear dynamics of the Navier--Stokes equations add further complexity to the analysis (see \S{}\ref{sec:cylinder_nonlinear}).

\subsection{Two-dimensional cylinder wake}

The wake of a cylinder is the most commonly used example of an oscillator-type shear flow.  The critical Reynolds number for onset of self-sustained vortex shedding, $Re_c=47$ according to the experiments by \cite{PMB87}, has been repeatedly recovered with high precision as the critical value for the onset of linear global instability \citep[see for instance][]{B06, SL07}. Furthermore, the nonlinear global instability of wakes is rather accurately predicted by local theory \citep{P02, C03}. The cylinder wake has been chosen to illustrate the sensitivity studies by \cite{GL07}, by \cite{MSJ08}, by \cite{LGP08} and by \cite{LB14}. The case at $Re=80$, well above the instability threshold, is considered in the present study. 

In the general notation (\ref{eqn:nonlinear}), the state vector $\bq=(\boldsymbol{u},p)$ now gathers the velocity vector and the pressure field, which together satisfy the incompressible Navier-Stokes equations, and the operators $\mathcal{B}$ and $\mathcal{N}$ are defined by 
\begin{eqnarray}\label{eqn:nonlinear_cyl}
\mathcal{B} = \left( 
\begin{tabular}{cc}
$\mathcal{I}$ & $0$ \\ 
$0$ & $0$ 
\end{tabular} \right) \;,\;\;
\mathcal{N}(\bq) = \left( 
\begin{tabular}{c}
$- (\bu \cdot \boldsymbol{\nabla} ) \bu - \boldsymbol{\nabla} p + \textit{Re}^{-1} \boldsymbol{\Delta} \bu $ \\ 
$\boldsymbol{\nabla}\cdot\bu$ 
\end{tabular} \right) .
\end{eqnarray}

\subsubsection{Linear instability of the cylinder wake}
\label{sec:cyl_lin_modes}

\begin{figure}
\centering
\begin{tabular}{ll}
(a) & (b)\\
\includegraphics[width=0.48\textwidth]{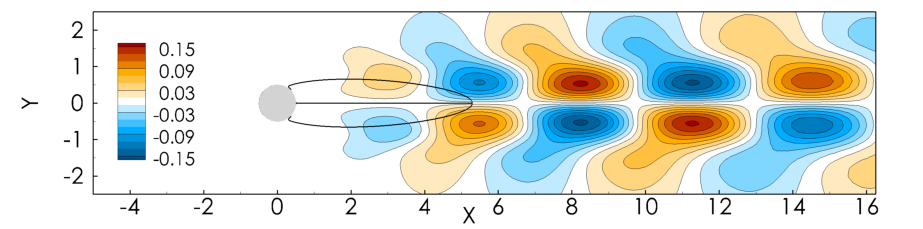} & \includegraphics[width=0.48\textwidth]{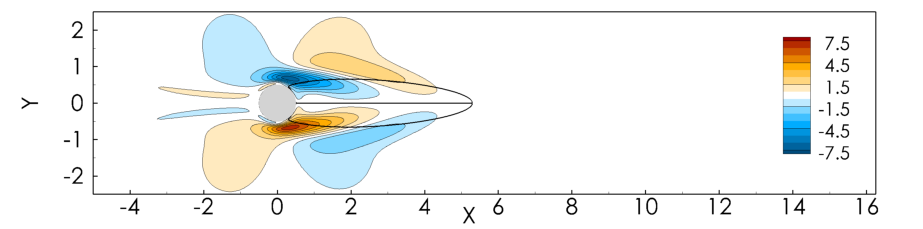}
\end{tabular}
\caption{Unstable linear eigenmode of the cylinder wake at $Re=80$. The real frequency is $\omega_r=0.785$, and the temporal growth rate is $\omega_i=0.102$. a) streamwise velocity perturbations $u$ of the direct eigenmode; b) adjoint streamwise velocity perturbations $u^\dagger$ of the associated adjoint mode.The black line indicates the recirculation region in the steady flow.}
\label{fig:cyl_lin_u}
\end{figure}

The cylinder wake problem permits
a non-trivial steady solution $\bq_b=(\bu_b,p_b)$, satisfying $\mathcal{N}(\bq_b)=\boldsymbol{0}$, which will serve as a base flow. Linear perturbations $\bq'=(\bu',p')$ developing on this base flow  are governed by the linear equations (\ref{eqn:linear}) with
\begin{equation}\label{eqn:L_cyl}
\mathcal{L}_{\bq_b} \bq' = \left( 
\begin{tabular}{c}
$- (\bu_b \cdot \boldsymbol{\nabla} )\bu' - ( \bu' \cdot \boldsymbol{\nabla} ) \bu_b + \textit{Re}^{-1} \boldsymbol{\Delta}\bu' -\boldsymbol{\nabla}p'$ \\ 
$\boldsymbol{\nabla} \cdot \bu' $ 
\end{tabular} \right) .
\end{equation}
Eigenmodes of this linear operator are obtained numerically as described in \cite{SL07}. Only one unstable mode is found for $\textit{Re}=80$, with an eigenvalue $\lambda=\sigma-\i\omega=0.1018-0.7852\i$ . The streamwise velocity of the real part of this mode is shown in figure \ref{fig:cyl_lin_u}({\em a}). The black line represents the stagnation-point streamline of the base flow, demarcating the recirculation region. The  streamwise velocity $u^\dagger$ of the associated adjoint eigenmode is displayed in figure \ref{fig:cyl_lin_u}({\em b}). As discussed in detail by \cite{GL07}, \cite{MSJ08} and others, the direct and adjoint eigenmode are localized upstream and downstream of the recirculation region, respectively. 

\subsubsection{Nonlinear instability of the cylinder wake}
\label{sec:cyl_nonlin_modes}

\begin{figure}
\centering
\begin{tabular}{ll}
(a) & (b)\\
\includegraphics[width=0.48\textwidth]{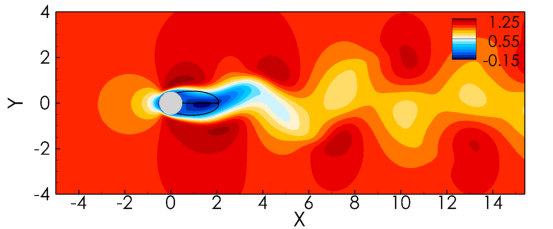} & \includegraphics[width=0.48\textwidth]{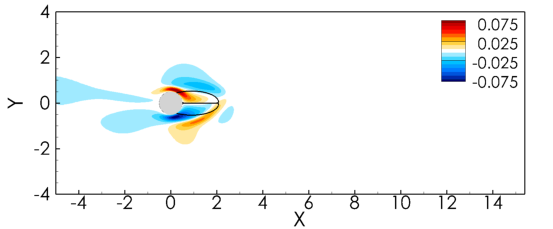}
\end{tabular}
\caption{Nonlinear global mode of the cylinder wake at $Re=80$: a) total streamwise velocity $U+u$; b) adjoint streamwise velocity perturbations $u^\dagger$ of the associated adjoint Floquet mode. Both snapshots are taken at the same instant. The black line indicates the recirculation region in the mean flow.}
\label{fig:cyl_nonlin_u}
\end{figure}

As the growing linear eigenmode reaches finite amplitude levels, the nonlinear terms  become significant  and the cylinder wake  settles into a saturated  periodically oscillating state, the B\'enard--von K\'arm\'an vortex street. This time-periodic solution $\bq_{0}(\bx,t)$ is obtained numerically by time-marching the nonlinear equations with a semi-implicit second-order temporal discretisation, and a spatial discretisation identical to the one used for the eigenvalue problem. At each temporal iteration, an unsteady Stokes problem is solved with a preconditionned Uzawa algorithm \citep[see][]{CSS86} implemented in the FreeFem++ software.  At $Re=80$, the global frequency is found to be $\omega_g=0.9957$, which is to be compared to the frequency of the linear eigenmode, $\omega=0.7852$. The nonlinear correction to the global frequency is significant in this supercritical setting. A snapshot of the total streamwise velocity of the nonlinear global mode is shown in figure \ref{fig:cyl_nonlin_u}({\em a}), where the black line now represents the stagnation-point streamline contour of the time-averaged flow. In the same way as discussed in \S{}\ref{sec:nonlinGL} for the nonlinear Ginzburg--Landau equation, see (\ref{eqn:floquet_adjoint}, \ref{eqn:GL_adjoint}), an adjoint mode associated with the neutral Floquet mode of the nonlinear periodic state can be obtained by backward time-stepping of the adjoint tangential equation. Its streamwise velocity component $u^\dagger$ is shown in figure \ref{fig:cyl_nonlin_u}({\em b}), taken at the same instant as  the direct flow field in  figure \ref{fig:cyl_nonlin_u}({\em a}).

\section{Active flow regions in linear global modes}

\subsection{Sensitivity of the eigenvalue}
\label{sec:linear_sensitivity}

The sensitivity of an eigenvalue measures how this value varies in response to changes of the operator. In the present context, only \emph{linear} sensitivities are considered, i.e. all variations are assumed to be infinitesimally small. If the original linear equation \eqref{eqn:linear} has eigenmodes that satisfy $\lambda_j \mathcal{B}\bphi _j = \mathcal{L}\bphi _j$, a small perturbation $\eps\delta\mathcal{L}$ of the operator leads to a perturbed eigenvalue problem,
\begin{equation}
(\lambda_j+\eps\del\lambda_j) \mathcal{B} (\bphi _j + \eps\del\bphi _j) = (\mathcal{L} + \eps\del\mathcal{L})(\bphi _j + \eps\del\bphi _j) \, ,
\end{equation}
which at order $\eps$ can be rearranged to give
\begin{equation}
(\lambda _j \mathcal{B} - \mathcal{L} ) \, \del\bphi _j = -\del\lambda_j \mathcal{B} \bphi _j + \del\mathcal{L} \, \bphi _j \, .
\end{equation}
The Fredholm alternative states that this inhomogeneous problem in $\del\bphi _j$ has a solution if and only if the right-hand side term is orthogonal to the nullspace of the left-hand side operator, i.e.~ $\langle \bphi _j^{\dag} , -\del\lambda_j \mathcal{B} \bphi _j + \del\mathcal{L} \, \bphi _j \rangle =0 $. If the direct and adjoint modes are normalized such that $\langle \bphi _j^{\dag} \, ,\, \mathcal{B} \bphi _j \rangle=1$, this condition leads to
\begin{eqnarray}\label{eqn:eigdrift}
\del\lambda_j = \langle \bphi _j^{\dag} \, ,\, \delta \mathcal{L} \, \bphi _j \rangle \, .
\end{eqnarray}
On the basis of \eqref{eqn:eigdrift}, \cite{GL07} consider spatially localized perturbations of the operator,
\begin{eqnarray}\label{eqn:giannettimodif}
\delta \mathcal{L} = \delta(\bx-\bx_{0}) \, \mathcal{C}_{0} \, ,
\end{eqnarray}
where $\mathcal{C}_0$ represents some artificially added coupling between the various flow variables at the location  $\bx_{0}$. If $\bphi _j(\bx_{0})$ and $\bphi _j^\dagger(\bx_{0})$ are understood to be vectors containing the $n$ flow variable values at $\bx_{0}$ ($n=1$ for the Ginzburg--Landau equation and $n=3$ for the cylinder wake), then $\mathcal{C}_{0}$ is represented by an $n\times n$ matrix, $\mathsfbi{C}_0$, and the eigenvalue variation is obtained from (\ref{eqn:eigdrift}) as
\begin{equation}\label{eqn:eigdrift2}
\del\lambda_j\big|_{\bx_0}= \bphi _j^{\dagger \star}(\bx_{0}) \cdot \mathsfbi{C}_0 \cdot  \bphi _j(\bx_{0}) \, . 
\end{equation}
Taking the norm of $\mathsfbi{C}_0$ to be unity without loss of generality, application of the Cauchy--Schwarz theorem yields an upper bound for the modulus of the eigenvalue variation, induced by an operator variation at $\bx_{0}$ \citep{GL07}:
\begin{equation}
|\delta\lambda_j |_{\bx_0} \le  \| \bphi _j^\dagger(\bx_{0}) \| \,  \| \bphi _j(\bx_{0}) \| \, .
\end{equation}

\cite{MSJ08} define the operator variation in \eqref{eqn:eigdrift} specifically as being due to variations $\del U$ of the base flow, $\del\mathcal{L} = (\nabla_U \mathcal{L}) \, \del U$. This definition allows to quantify how a given small modification of the base flow, localized or distributed, alters the frequency $\omega$ and the growth rate $\sigma$. Both approaches represent the mathematical formulation of a well-posed question, based on different interpretations of what constitutes a `wavemaker': in the case of \cite{GL07}, it is the localized `internal feedback' between perturbations, whereas in the case of \cite{MSJ08} it is the feeding of perturbation growth on base flow energy. In both approaches, the answer is sought by probing the system with \emph{exogenous} modifications of the operator structure.

\subsection{Endogeneity analysis of linear global modes}
\label{sec:endo_lin}

With the question in mind how a localized region in the flow contributes to the global dynamics, we note that the admittance of any arbitrary operator $\mathcal{C}_0$ in \eqref{eqn:giannettimodif} may be too general for the purpose of identifying the specific interactions that are inherent in the linear Navier--Stokes operator. Retaining the idea of considering the sensitivity of the eigenvalue with respect to \emph{localized} changes of the operator, we stipulate that those changes preserve the local \emph{structure} of the operator. This naturally leads to choosing
\begin{equation}
\delta \mathcal{L} = \delta(\bx-\bx_{0}) \, \mathcal{L} \, .
\end{equation}
The variation of the operator at $\bx_0$ is chosen to be proportional to the original operator itself in that same location. The sensitivity with respect to such variations quantifies directly how much the eigendynamics in a given point in space contribute to the frequency and to the growth rate; it is therefore suitable for an investigation of the \emph{endogenous} global dynamics.  We call this specific sensitivity
\begin{equation}
\label{eqn:endo_linear}
E(\bx) = \bphi _j^{\dag \star}(\bx) \cdot \left( \mathcal{L} \bphi _j \right)(\bx)
\end{equation}
the \emph{endogeneity} of the eigenmode $(\lambda_j, \bphi _j)$. Its computation is straightforward if the direct and adjoint eigenmodes as well as the operator are available.  The dot-product in \eqref{eqn:endo_linear} again only denotes the scalar multiplication of two vectors containing the various state variables in one point in space.

An essential property of the endogeneity is that its integral over $\bx$ is equal to the eigenvalue $\lambda _j$:
\begin{equation}
\int_\Omega E(\bx) \text{d}\bx = \int_\Omega \bphi _j^{\dag \star} \cdot \left( \mathcal{L} \bphi _j \right) \text{d}\bx = \langle \bphi _j^{\dag} \, , \,  \mathcal{L} \bphi _j \rangle = \langle \bphi _j^{\dag} \, , \, \lambda_j \mathcal{B} \bphi _j \rangle = \lambda_j \, .
\end{equation}
It is important to note that, while any spatial distribution can be normalised to yield any integral scalar value, the endogeneity is the unique quantity that represents \emph{local} contributions to the eigenvalue. For instance, the endogeneity allows to exclude any point in space from the integration, and the result reflects how the eigenvalue is altered due to the missing local contribution.

An endogeneity-based analysis clearly distinguishes between the promotion of unstable growth, contained in the real part  of $E(\bx)$, and the frequency selection, given by the negative imaginary part. For the sake of clarity, let
\begin{equation} 
\label{eqn:endo_lambda}
E_\sigma(\bx) = \Re\left[ E(\bx)\right] \text{~and~}  E_\omega(\bx) = -\Im\left[ E(\bx)\right]
\end{equation}
be defined, such that $E(\bx) = E_\sigma(\bx) - \i E_\omega(\bx)$, analogous to $\lambda=\sigma - \i\omega$.
The distinction between these two components is of great importance for a physical discussion, and the following examples will show that $E_\sigma$ and $E_\omega$ in general present quite different spatial structures. Furthermore, with the definition \eqref{eqn:endo_linear} it is straightforward to decompose the operator $\mathcal{L}$, for instance into convection, diffusion and other terms, and to examine the individual contributions of these separate parts. Such a decomposition will be discussed in \S{}\ref{sec:cylinder_linear} for the cylinder wake.


\subsection{Example 1: linear Ginzburg--Landau equation}
\label{sec:GL_linear}

The endogeneity formalism is first applied to the global instability modes shown in figure \ref{fig:eigenfunctions_linGL}, with the parameters as given in \S{} \ref{sec:linGL}. Choosing $\mu_0=\mu_c$, as defined by (\ref{eqn:mu_c}), the system is marginally unstable in a global sense. 
The endogeneity is found by multiplying $\phi^{\dagger *}$ with $\mathcal{L}\phi$ in every point $x$, where the eigenfunctions $\phi(x)$ and $\phi^\dagger(x)$ are given analytically by (\ref{eqn:GL_lin_direct}, \ref{eqn:GL_lin_adjoint}), and the operator $\mathcal{L}$ is written out in (\ref{eqn:GL_lin}).
\begin{figure}
\centering
\includegraphics[width=\textwidth]{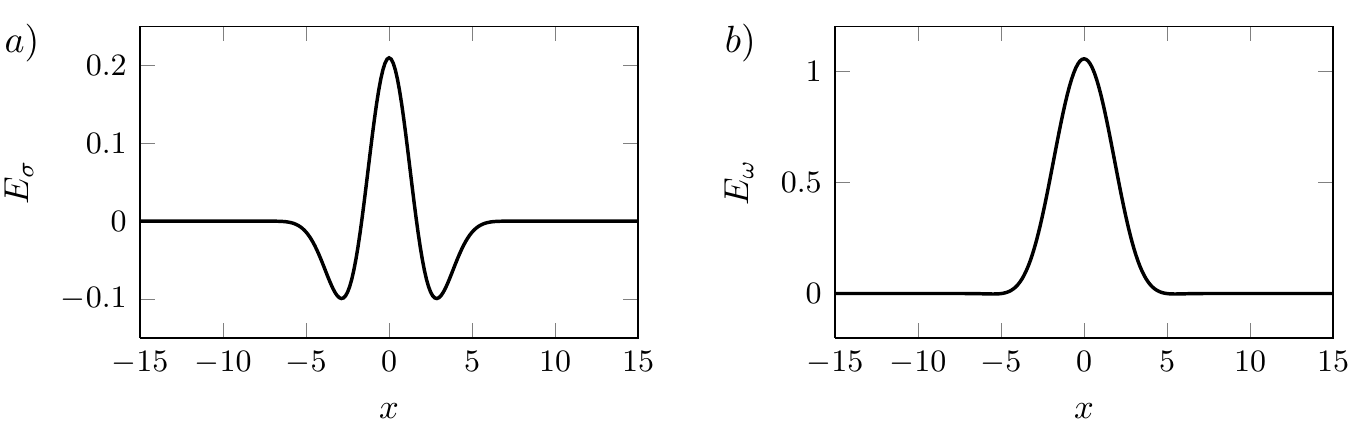}
\caption{Endogeneity distribution of the leading linear Ginzburg--Landau eigenmode. (a) real part $E_\sigma$, related to growth rate, ($b$) negative imaginary part $E_\omega$, related to frequency. }
\label{fig:GL_endo_linear}
\end{figure}

Both parts of the endogeneity, $E_\sigma$ and $E_\omega$, are shown in figure \ref{fig:GL_endo_linear}: both are even functions, with their maximum values at $x=0$.  The integral of $E(x)$ is exactly equal to the associated eigenvalue, $\lambda = -4.398\i$. In the $E_\sigma(x)$ distribution, shown in figure \ref{fig:GL_endo_linear}(a), negative and positive regions exactly counterbalance each other, totalling a zero growth rate.  The largest contribution to both the frequency selection and the growth rate stems from the region around $x=0$, not from regions where the magnitude of either the direct or adjoint eigenmodes is large (compare to figure \ref{fig:eigenfunctions_linGL}).

This result is compared to the saddle point criterion given by \cite{CHR91}. In their model, the `wavemaker' location is defined by a saddle point of the local absolute frequency $\omega_0$ in the complex $x$ plane, $\partial_x \omega_0 = 0$. In the present case, one finds $\omega_0(x) = i\mu(x) - iU^2/4\gamma$, and therefore $\partial_x\omega_0=i \partial_x \mu$, with a single saddle point precisely at $x=0$.
The `wavemaker' location according to the criterion of \cite{CHR91} is identical with the \emph{maximally endogenous} location in this example.  But whereas their WKBJ-based criterion identifies a singular location, the endogeneity quantifies contributions to the growth rate and the frequency from any point in the domain, thereby characterizing a distributed `wavemaker'.

\subsection{Example 2: unstable  linear  global mode of the cylinder wake}
\label{sec:cylinder_linear}

\begin{figure}
\centering
\begin{tabular}{ll}
(a) & (b)\\
\includegraphics[width=0.48\textwidth]{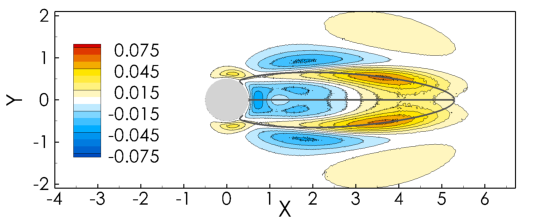} & \includegraphics[width=0.48\textwidth]{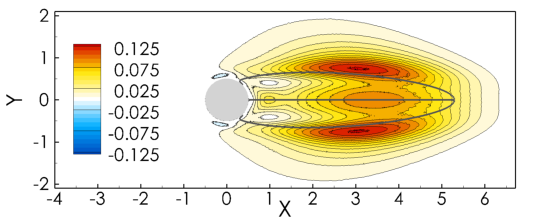}
\end{tabular}
\caption{Endogeneity distribution of the linear cylinder wake instability. {\em a}) $E_\sigma(\bx)$,  related to growth rate; {\em b}) $E_\omega(\bx)$,  related to frequency. The spatial integral of $E_\sigma$ equals the growth rate $\sigma=0.102$, the spatial integral of $E_\omega$ equals the frequency $\omega=0.785$.} 
\label{fig:endo_cyl_lin}
\end{figure}

The endogeneity of the unstable linear eigenmode $\bphi$ of the cylinder wake, displayed in figure \ref{fig:cyl_lin_u}(a), is now examined. It is readily computed by point-wise multiplication of the complex conjugate of the adjoint eigenmode $\bphi^{\dag}$, displayed in figure \ref{fig:cyl_lin_u}(b),  with $\mathcal{L} \bphi$ where the definition  of $\mathcal{L}$ is given in (\ref{eqn:L_cyl}). As the divergence of perturbation velocity is zero in every point, the continuity equation along with the adjoint pressure $p^\dag$ vanishes from the endogeneity definition, and one is left with
\begin{equation}
E(\bx) = -\bu^{\dag *}\cdot[(\bu_b\cdot\nabla)\bu] - \bu^{\dag *}\cdot[(\bu\cdot\nabla)\bu_b]  -\bu^{\dag *}\cdot\nabla p + Re^{-1} \, \bu^{\dag *}\cdot \Delta \bu \, ,
\label{eqn:endo_lin_cyl}
\end{equation}
where again it is understood that the left-hand side is to be evaluated in every point $\bx$, such that all vectors only contain two scalar elements ($x$- and $y$-components). The endogeneity of the linear cylinder-wake instability mode is displayed in figure \ref{fig:endo_cyl_lin}. The distribution of $E_\sigma(\bx)$, depicted in figure \ref{fig:endo_cyl_lin}(a), shows where the temporal growth rate is generated, whereas $E_\omega(\bx)$, shown in figure \ref{fig:endo_cyl_lin}(b), indicates how the various flow regions influence the global frequency selection. 

A few general conclusions can be inferred from figure \ref{fig:endo_cyl_lin}. Mainly, it is observed that the endogeneity (real and imaginary parts) is concentrated around the shear layers of the separation region, delimited in the figures by black lines. The frequency selection is clearly concentrated in two symmetric maximum locations. This part of the endogeneity resembles the quantity displayed by \cite{GL07} in their figure 17. The distribution of $E_\sigma(\bx)$ however bears a more faceted structure. Some regions are positive, contributing to global instability, others are negative, thus stabilising the eigenmode. The entire flow downstream of the separation region has practically no influence on frequency and growth rate, consistent with the conclusions of \cite{GL07} and \cite{MSJ08}.

\begin{figure}
\begin{tabular}{ll}
(a) & (b)\\
\includegraphics[width=0.48\textwidth]{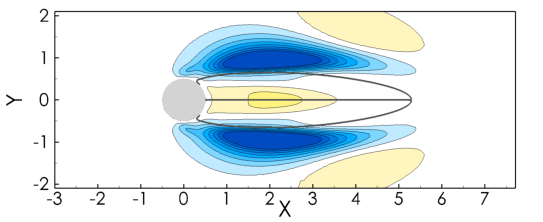} & \includegraphics[width=0.48\textwidth]{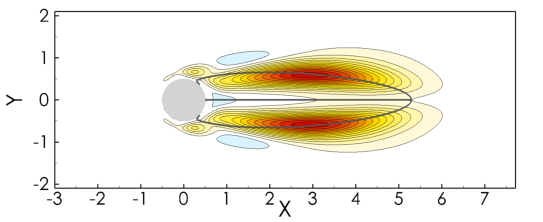} \\
(c) & (d)\\
\includegraphics[width=0.48\textwidth]{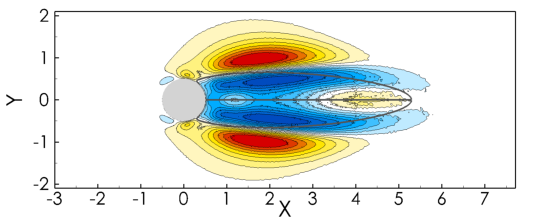} & \includegraphics[width=0.48\textwidth]{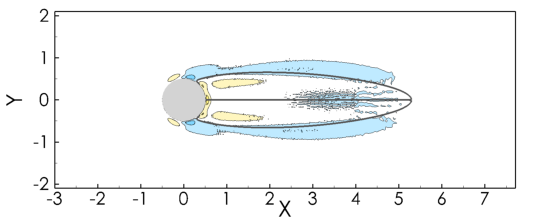} 
\end{tabular}
\caption{Real parts of the individual endogeneity terms in (\ref{eqn:endo_lin_cyl}) (from left to right). {\em a}) Convection by the base flow; {\em b}) production through base flow shear; {\em c}) pressure forces; {\em d}) diffusion. The sum of these contributions gives $E_\sigma(\bx)$, shown in figure \ref{fig:endo_cyl_lin}{\em a}. The same color scale is used here.}
\label{fig:cyl_lin_decomposition}
\end{figure}

A more insightful analysis of physical mechanisms may be based on a decomposition of the endogeneity into contributions from the various left-hand side terms in (\ref{eqn:endo_lin_cyl}). In the given order, these terms account for the effects of base flow convection; production through base flow shear; pressure forces and diffusion. Their spatial distributions (real parts only, reflecting contributions to the growth rate) are shown in figure \ref{fig:cyl_lin_decomposition}. The effect of convection by the base flow is dominantly stabilising (figure \ref{fig:cyl_lin_decomposition}\emph{a}). In the outer vicinity of the separation bubble, the downstream convection of perturbations counteracts their capacity of \emph{in situ} growth, and renders the instability more convective. Inside the recirculation region, the upstream convection has the opposite effect. 
The production term (figure \ref{fig:cyl_lin_decomposition}\emph{b}) provides the principal source of global unstable growth.

The workings of pressure forces are not as obvious to interpret in physical terms, and their integrated net contribution to $E_\sigma$ and $E_\omega$ is exactly zero in an incompressible setting. Since the adjoint velocity field is divergence-free, $\nabla\cdot\bu^\dag = 0$, it is easily found that 
\begin{equation}
\int_\Omega \bu^\dag \cdot \nabla p \text{~d}\bx = -\int_\Omega( \nabla\cdot\bu^\dag )\, p \text{~d}\bx =0
\end{equation}
Yet this term contributes strongly to the local values of the overall endogeneity. Tentatively, it may be argued that the role of the perturbation pressure gradient is to enforce the continuity condition, thereby causing a perturbation volume flux across the shear layer. As the instability perturbations tend to shorten the length of the separation bubble (manifest in the nonlinear mean flow), the stagnation-point streamline is forced toward the symmetry line, lessening the convective effect outside the bubble, and enhancing it inside.

\begin{figure}
\includegraphics[width=\textwidth]{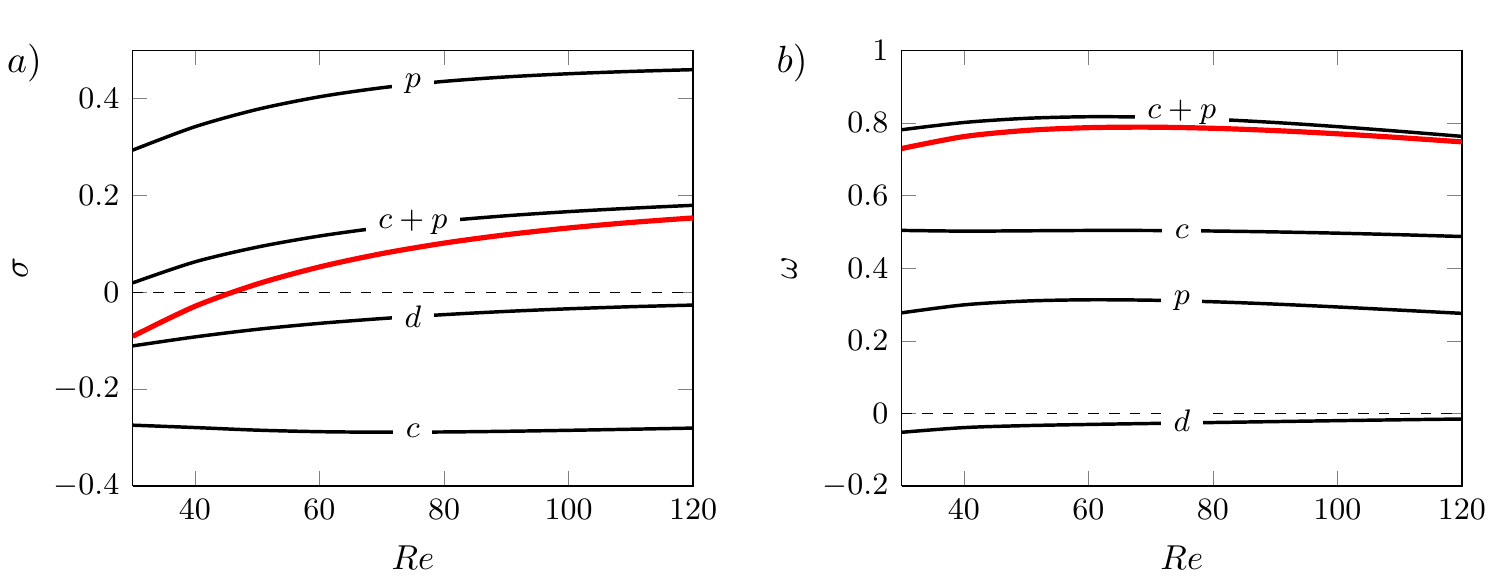} 
\caption{Net contributions of individual terms in the endogeneity definition (\ref{eqn:endo_lin_cyl}), 
({\em a}) to the linear growth rate, and ({\em b}) to the global frequency, as a function of the Reynolds number $\textit{Re}$. Convection (-$c$-), production (-$p$-) and diffusion
(-$d$-) contributions are shown, as well as the sum of convection and production ($c+p$) and the total values of $\sigma$ and $\omega$ (thick red lines). The pressure gradient contribution is zero for all Reynolds numbers. }
\label{fig:convproddiff}
\end{figure}

The diffusion term is globally stabilizing, although inside the separation bubble near the cylinder it provokes a weak destabilisation. This seems to be the consequence of viscous transport of perturbation velocity into the shear layer. However low the amplitude of the diffusive contribution, it is crucial for accurately determining the instability threshold. The Reynolds number is expected to have two distinct effects on the instability. First, the stabilizing effect of the perturbation diffusion should weaken when the Reynolds number is increased. This is confirmed in figure \ref{fig:convproddiff}(a), which shows the trends of the convection, production and diffusion contributions to the growth rate, as functions of the Reynolds number. The stabilizing diffusion effect lessens with increasing Reynolds number. Secondly, the steady base flow is influenced by the Reynolds number, affecting the instability mechanism via the production and convection terms. Individually, these contributions appear to be dominant, but with opposite effects on the growth rate. As  the Reynolds number is increased, the destabilisation by the production term overcomes the stabilisation by the convection term. The \emph{combined} effect of these competing terms, represented by a line marked $c+p$ in figure \ref{fig:convproddiff}(a), is then  comparable in strength to the diffusion effect. 
On the other hand, the role of the diffusion in the frequency selection is all but negligible, as seen in figure \ref{fig:convproddiff}(b). As for the growth rate, the dominant contributors are the production and convection terms. Both contribute here to increase the frequency. Their combined effect ($c+p$) gives a good prediction of the linear frequency (red line), especially at high Reynolds number.

\section{Active flow regions in nonlinear global modes}

\subsection{Sensitivity of the global frequency}

The general sensitivity formulation for nonlinear time-periodic oscillations of a bifurcated flow state adopted here is similar to the analysis by \cite{LGP08}, with some variations in the notation. Consider a nonlinear global mode $\bq_0(\bx,t)$, time-periodic solution of (\ref{eqn:nonlinear}), with fundamental frequency $\omega_g$. In order to make the frequency explicitly visible in the equation, the time variable is rescaled as $\tau=\omega_g t$, such that $\bq_0(\bx,\tau)$ is $2\pi$-periodic in $\tau$ and satisfies 
\begin{equation}
\omega_g \, \mathcal{B} \, \partial_\tau \bq_0 = \mathcal{N}(\bq_0)\, . 
\label{eqn:nonlinear_rescaled}
\end{equation}
Just as in the linear case of \S{}\ref{sec:linear_sensitivity}, small variations of the left-hand side operator cause variations of the solution, including its frequency:
\begin{equation}
(\omega_g + \eps\delta\omega_g) \, \mathcal{B} \, \p_\tau(\bq_0 + \eps\delta\bq_0) = (\mathcal{N}+\eps\delta\mathcal{N})(\bq_0 + \eps\delta\bq_0) \, .
\end{equation}
With the introduction of the tangential linear operator $\mathcal{L}_{\bq_0(t)}$ and its neutral Floquet mode $\bpsi_1 = \partial_{\tau} \bq_0$, defined in (\ref{eqn:tangential}, \ref{eqn:floquetmode}), variations are governed at order $\eps$ by the relation
\begin{equation}
(\omega_g\mathcal{B} - \mathcal{L}_{\bq_0(t)}) \, \delta\bpsi_1 =  -\delta\omega_g \, \mathcal{B} \, \bpsi_1 + \delta\mathcal{N}(\bq_0) \, ,
\end{equation}
The Fredholm alternative can be written out with the aid of the neutral adjoint Floquet mode $\bpsi_1^\dag$ (see \S{}\ref{sec:global_modes}), normalised to give $\{  \bpsi_1^\dag, \mathcal{B}  \bpsi_1\}=1$, which leads to
\begin{equation}
\delta\omega_g = \big\{ \bpsi_1^\dag, \delta\mathcal{N}(\bq_0) \big\} \, .
\label{eqn:sensitivity_nonlin}
\end{equation}
If $\bpsi_1^\dag$ is known, then the impact of any small operator variation $\delta\mathcal{N}$ on the global frequency can be immediately evaluated from (\ref{eqn:sensitivity_nonlin}). In practice, $\bpsi_1^\dag$ is obtained in the following way: first, the nonlinear equation is numerically integrated by time-stepping until the periodic nonlinear global mode regime is fully attained. Then the (linear) adjoint tangential equation is stepped backwards in time, starting from an arbitrary initialization, over as many cycles of the nonlinear global mode as necessary. During this backward-in-time integration, the adjoint solution converges asymptotically towards the sought-after neutral mode.

\subsection{Endogeneity analysis of nonlinear global modes} 
Analogously to the linear case in \S{ \ref{sec:endo_lin}, the endogeneity of a nonlinear global mode is defined by considering variations of the nonlinear operator that preserve its structure but are localized in space, and also in time,  since the nonlinear operator is time-dependent, 
\begin{equation}
\delta\mathcal{N} = \delta(\bx-\bx_0) \delta(\tau-\tau_0) \mathcal{N} \, .
\end{equation}
The inner product in (\ref{eqn:sensitivity_nonlin}) involves integration in $\bx$ as well as in $\tau$. Integration over the Dirac functions yields the expression for the influence of spatio-temporal variations in the operator on the frequency selection,
\begin{equation}
E_\omega(\bx,\tau)  = \Re \left[ \bpsi_1^{\dag *}(\bx,\tau) \cdot \mathcal{N}(\bq_0(\bx,\tau)) \right] .
\label{eqn:endo_nonlinear}
\end{equation}
%
Only the frequency selection in nonlinear global modes is considered at present, because it is assumed here that both quantities, in practice, are obtained from a flow solver in the form of real-valued variables. The scaled time $\tau$  denotes the temporal phase within an oscillation cycle. 

\subsection{Example 1: the nonlinear Ginzburg--Landau equation}
\label{sec:GL_nonlinear}

The nonlinear global mode and its neutral adjoint Floquet mode of the Ginzburg--Landau equation (figure \ref{fig:eigenfunctions_nonlinGL}) have been discussed in \S{}\ref{sec:nonlinGL}. The associated endogeneity, according to (\ref{eqn:endo_nonlinear}), is shown in figure \ref{fig:endo_nonlinGL}. It is noted that the endogeneity is independent of $\tau$ in the present case of the  Ginzburg--Landau equation.

\begin{figure}
\centering
\includegraphics[width=0.6\textwidth]{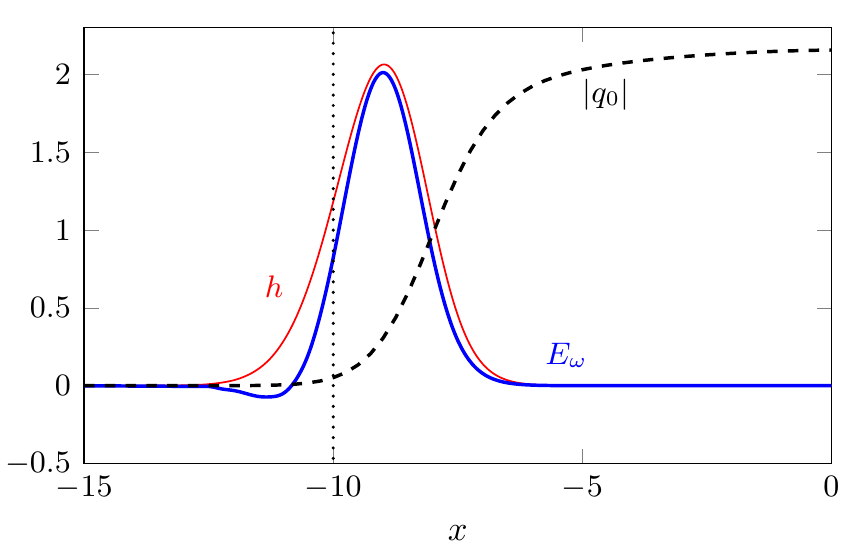}
\caption{Endogeneity $E_\omega$ (thick blue line) of the nonlinear global mode of the Ginzburg--Landau equation, as shown in figure \ref{fig:eigenfunctions_nonlinGL}, compared with the quantity $h(x)$ (thin red line), which is obtained with the formalism described by \cite{H15} according to (\ref{eqn:hwang}). The dotted line marks the location $x_{ca}$, and the dashed line traces the envelope shape of the nonlinear global mode.}
\label{fig:endo_nonlinGL}
\end{figure}

The $E_\omega$ distribution is nonzero in a narrow region around $x=-9$, and its integral in $x$ and over one period $0 \le\tau\le 2\pi$ is exactly equal to the global frequency, $\omega_g=3.70$. According to \cite{PHCC98}, the global frequency of a nonlinear Ginzburg--Landau system with slowly varying coefficients is selected at the location $x_{ca}$, where the local instability changes from upstream convective to downstream absolute. In the limit of marginal global instability, the global frequency is then predicted to correspond to the absolute frequency at $x_{ca}$. In the present example, one finds $x_{ca}=-10$, and the absolute frequency at this location is $\omega_0(x_{ca}) = 4.50$. The chosen parameter configuration (see \S{}\ref{sec:GL}) is strongly globally unstable, and the parameter $\mu$ varies significantly in $x$ around $x_{ca}$, therefore the present case does not respect the limiting assumptions of \cite{PHCC98}, and the frequency prediction is quite inaccurate as a result. The endogeneity however provides a clear and accurate picture of the frequency selection process. The maximum contribution to the global frequency is found at $x=-9$, indeed not far from the `wavemaker' location $x_{ca}$ as defined by  \cite{PHCC98}.


The present results may be compared to the structural sensitivity of nonlinear global modes as defined by \cite{H15}, who adapted the formulation of \cite{LGP08} to analyze nonlinear global modes of the Ginzburg--Landau equation. According to \cite{H15}, worst-case variations of the global frequency due to added `closed-loop perturbations' (synonymous to `internal feedback'),  written in the notation of this paper, are characterized by the spatial distribution
\begin{equation}
h(x) = \left| \frac{q_0^*(x)\, \psi^\dag_1(x)}{N} \right| ,  \text{~~with~~} N=\int_0^T\!\!\! \int_{-\infty}^\infty ( \psi_1^{\dag *}\, q_0 - \psi_1^{\dag}\, q_0^*) \, \text{d}x \, \text{d}t\, .
\label{eqn:hwang}
\end{equation}
Figure \ref{fig:endo_nonlinGL} compares this distribution, multiplied with $2\omega_g$ for consistent scaling, with the endogeneity $E_\omega$. The two curves are very similar, and in particular the position of their maxima is identical. Again it is pointed out that the analysis of \cite{H15}, in contrast to the endogeneity formalism, considers modifications of the operator that do not preserve its original structure.

In conclusion, just as in the linear analysis of \S{}\ref{sec:endo_lin}, the results obtained from the nonlinear endogeneity analysis are consistent with the `wavemaker' definition from classical asymptotic theory, but they go further in a quantitative description of the dynamics, because the endogeneity fully accounts for the effects of non-parallelism and supercriticality. It is also consistent with recent formulations that describe the structural sensitivity, but it reveals the endogenous dynamics that are specific to the operator under consideration. 

\subsection{ Example 2:  Nonlinear frequency selection in the cylinder wake}\label{sec:cylinder_nonlinear}

The endogeneity of the time-periodic flows developing in the wake of the circular cylinder, described in \S{}\ref{sec:cyl_nonlin_modes}, is explicitly obtained as
\begin{equation}\label{eqn:endo_nonlin_cyl}
E_\omega(\bx,\tau) = \bu^\dag_1\cdot \left[ -(\bu_0\cdot\nabla)\bu_0 - \nabla p_0 + Re^{-1} \Delta \bu_0 \right] ,
\end{equation}
where $\bq_0(\bx,\tau)=(\bu_0,p_0)$ is the $2\pi$-periodic nonlinear solution and $\bu^{\dagger}_1$ is the velocity component of the $2\pi$-periodic solution of the linear adjoint equation.
\begin{figure}
\centering 
\includegraphics[width=0.7\textwidth]{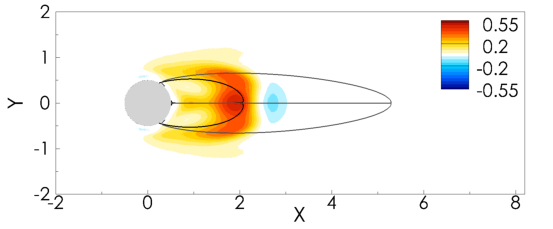}
\caption{Time-integrated endogeneity $E_\omega(\bx)$ of the nonlinear global mode in the cylinder wake at $\textit{Re}=80$. The nonlinear frequency is $\omega_{g}=0.995$. The black (resp. grey) lines indicate the recirculation region in the time-averaged mean flow (resp. base flow).}
\label{fig:endocylnonlin}
\end{figure}

The spatio-temporal endogeneity (\ref{eqn:endo_nonlin_cyl}) is integrated over one oscillation cycle, and the result,  shown  in figure \ref{fig:endocylnonlin},  demonstrates  that the endogenous region  resembles  a front, localized around $x=2$. Similarly to   the earlier observations in the context of  the Ginzburg-Landau equation, the endogenous region of the nonlinear global mode  in the cylinder wake is located further upstream than that of its linear counterpart  (compare to  figure \ref{fig:endo_cyl_lin}b). The separation line of the recirculation regions both in the base flow and in the time-averaged mean flow
are depicted by grey and black lines,  respectively, in figure \ref{fig:endocylnonlin}. Interestingly, the endogenous region of the nonlinear global mode  appears to be supported by  the recirculation region in the mean flow, as opposed to the base flow. This  observation  suggests  that further investigation of  the nonlinear dynamics   should focus on the analysis of  fluctuations around the mean flow state.

To  this  aim,  the nonlinear global mode state is  first decomposed as  $\bq_0 = \boldsymbol{\bar{q}}+\bq' $ into a time-averaged  component  $\boldsymbol{\bar{q}}=(\boldsymbol{\bar{u}},\bar{p})^{T}$ and a  zero-mean   fluctuation   component  $\bq'=(\bu',p')^{T}$. Introducing  this  decomposition into the Navier-Stokes operator (\ref{eqn:nonlinear_cyl}), and time-averaging over one oscillation cycle, one obtains the steady nonlinear equations  that govern  the mean flow,
\begin{align}
\label{eqn:mean}
(\boldsymbol{\bar{u}}\cdot\nabla)\boldsymbol{\bar{u}} + \nabla \bar{p} - Re^{-1} \Delta \boldsymbol{\bar{u}}   &= - \overline{ (\boldsymbol{u'} \cdot \nabla) \boldsymbol{u'} } , \\
\nabla \cdot \boldsymbol{\bar{u}} &= 0 \, .
\end{align}
By subtraction, the unsteady nonlinear equations for the fluctuations are obtained as
\begin{align}
\label{eqn:fluct}
\omega_{g} \, \partial_{\tau} \boldsymbol{u'} + (\boldsymbol{\bar{u}}\cdot\nabla)\boldsymbol{u'} + (\boldsymbol{u'}\cdot\nabla)\boldsymbol{\bar{u}} + \nabla p' - Re^{-1} \Delta \boldsymbol{u'}
&= - (\boldsymbol{u'} \cdot \boldsymbol{\nabla}) \boldsymbol{u'}  + \overline{ (\boldsymbol{u'} \cdot \boldsymbol{\nabla}) \boldsymbol{u'} } ,\\
\nabla \cdot \boldsymbol{u}' &= 0 \, .
\end{align}
The right-hand side forcing terms both in (\ref{eqn:mean}) and in (\ref{eqn:fluct}) arise from the nonlinear interaction of fluctuations. The temporal mean of $(\boldsymbol{u'} \cdot \boldsymbol{\nabla}) \boldsymbol{u'}$ forces the mean flow, whereas its zero-mean fluctuation forces the flow fluctuations. 
At small amplitudes of $\boldsymbol{u'}$, these terms are negligible, such that the steady and unsteady equations reduce to the base flow and linear perturbation equations. By contrast, when a linear eigenmode experiences exponential growth and reaches finite amplitude levels, the effect of the forcing terms in (\ref{eqn:mean}) and (\ref{eqn:fluct}) will become significant. The right-hand side in (\ref{eqn:mean}) drives the base flow towards the mean flow. The right-hand side in (\ref{eqn:fluct}) modifies the dynamics of the fundamental oscillations at frequency $\omega_{g}$, and it generates harmonic components at integer multiples of $\omega_g$. Thus the effect of nonlinearity on the frequency selection can be attributed to two distinct origins. The first is a nonlinear deformation of the mean flow, which in turn modifies the left-hand side linear operator in (\ref{eqn:fluct}). The second one is the nonlinear interaction of fluctuation harmonics, induced by the right-hand side forcing term in (\ref{eqn:fluct}).

The endogeneity definition (\ref{eqn:endo_nonlin_cyl}) is now expanded in terms of mean flow and fluctuation components. With this decomposition, as developed above, one obtains
\begin{eqnarray}
E_\omega(\bx,\tau)  
&=& \bu^\dag \cdot \Big[ -(\boldsymbol{\bar{u}}\cdot\nabla)\boldsymbol{\bar{u}} -\nabla \bar{p} + Re^{-1} \Delta \boldsymbol{\bar{u}} - \overline{(\bu'\cdot\nabla)\bu'}   \Big] \\
&+& \bu^{\dag} \cdot \Big[ -(\overline{\bu}\cdot\nabla)\bu' - (\bu'\cdot\nabla)\overline{\bu} -\nabla p' + Re^{-1} \Delta \bu' \Big]  \nonumber \\
&+& \bu^\dag \cdot \Big[ -(\bu'\cdot\nabla)\bu' + \overline{(\bu'\cdot\nabla)\bu'}   \Big] \, . \nonumber 
\end{eqnarray}
The first line in the above equation vanishes because the expression in square brackets is the momentum equation for the mean flow (\ref{eqn:mean}. By further splitting the adjoint velocity into mean  and fluctuation  components, denoted $\overline{\boldsymbol{u}}^{\dag}$ and ${\boldsymbol{u}^{\dag}}'$ respectively, the time-integrated endogeneity simplifies to
\begin{equation}
\int_{0}^{2 \pi} E_\omega(\bx,\tau) \; d\tau 
= \int_{0}^{2 \pi} E_{m}(\bx,\tau) \; d\tau  + \int_{0}^{2 \pi} E_{h}(\bx,\tau) \; d\tau \, , \label{eqn:endo_decomp} 
\end{equation}
with
\begin{align}E_{m}(\bx,\tau) &=  {\boldsymbol{u}^{\dag}}' \cdot \left( -(\overline{\bu}\cdot\nabla)\bu' - (\bu'\cdot\nabla)\overline{\bu} -\nabla p' + Re^{-1} \Delta \bu' \right) \, , \label{eqn:endo_decomp_m} \\
 E_{h}(\bx,\tau) &= {\boldsymbol{u}^{\dag}}' \cdot \left( -(\bu'\cdot\nabla)\bu' + \overline{(\bu'\cdot\nabla)\bu'}   \right) \, ,\label{eqn:endo_decomp_h} 
\end{align}
All contributions involving the mean 
adjoint velocity vanish. The component $E_{m}(\boldsymbol{x},\tau)$ highlights spatial regions where the quasi-linear dynamics around the mean flow contribute to the frequency selection, whereas
$E_{h}(\boldsymbol{x},\tau)$ identifies regions where the interaction of harmonic components influences the global frequency.  

\begin{figure}
\begin{tabular}{ll}
(a) & (b) \\
\includegraphics[scale=0.5]{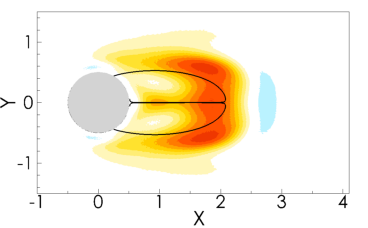} &
\includegraphics[scale=0.5]{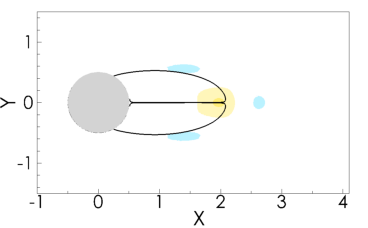}
\end{tabular}
\caption{Contributions to the time-integrated endogeneity of the nonlinear global mode. (a) Quasi-linear dynamics around the mean flow defined by (\ref{eqn:endo_decomp_m}) and (b) interaction between high harmonical components defined by (\ref{eqn:endo_decomp_m}). The color bar shown in figure \ref{fig:endocylnonlin} is also used here. The black line indicates the recirculating flow region in the mean flow. $\textit{Re}=80$.}  
\label{fig:endo_decompmean}
\end{figure}
The time-integrated components $E_m$ and $E_h$, for the nonlinear global mode in the cylinder wake at $\textit{Re}=80$, are shown in figure  \ref{fig:endo_decompmean}.  The sum of these two contributions, according to (\ref{eqn:endo_decomp}), gives the time-integrated frequency endogeneity displayed in figure \ref{fig:endo_decompmean}. In the present configuration, the frequency selection process is clearly dominated by the quasi-linear dynamics around the mean flow, whereas harmonic interactions  contributes only very weakly. The total contributions
\begin{equation}
\label{eqn:omega_decomp}
\omega_{m} = \int_{\Omega} \int_{0}^{2 \pi} E_{m}(\bx,\tau) \; d\tau \; d \boldsymbol{x} \, , \quad
\omega_{h} = \int_{\Omega} \int_{0}^{2 \pi} E_{h}(\bx,\tau) \; d\tau \; d \boldsymbol{x} \, ,
\end{equation}
are reported in table \ref{tab:meanfreq} for various values of the Reynolds number.  For all values of the Reynolds number investigated in this study, the contribution $\omega_{m}$  of the quasi-linear mean flow dynamics to the global frequency is greater than $97 \%$.
  
\begin{table}
\centering
\begin{tabular}{rrrcr}
$\textit{Re}$ \vline & $\omega$ \vline & $\omega_{g}$ \vline  & $\omega_{m}(\%)$ & $\omega_{h}(\%)$   \\
\hline
$50$ \vline & $0.7676$ \vline & $0.8159$ \vline & $0.8087 \, (99.1)$ & $0.0072 \, (0.9)$  \\
$75$ \vline & $0.7742$ \vline & $0.9726$ \vline & $0.9543 \, (98.1)$ & $0.0182 \, (1.9) $ \\
$80$ \vline & $0.7852$ \vline & $0.9957$ \vline & $0.9768 (98.1)$ & $0.0189 (1.9)$  \\
$100$ \vline & $0.7553$ \vline & $1.0703$ \vline &  $1.0493 \, (98.0)$ & $0.0210 \, (2.0)$  \\
$125$ \vline & $0.7246$ \vline & $1.1382$ \vline  & $1.1127 \, (97.7) $ & $0.0255 \, (2.3)$  \\
$150$ \vline & $0.6896$ \vline & $1.1922$ \vline & $1.1596 \, (97.2) $ & $0.0325 \, (2.8)$ 
\end{tabular}
\caption{ Endogeneity-based decomposition of the nonlinear global mode frequency, for various values of the Reynolds number $\textit{Re}$.
Linear eigenmode frequency $\omega$ of the base flow;  nonlinear global mode frequency
$\omega_{g}$; contributions to $\omega_g$ from quasi-linear dynamics in the mean flow ($\omega_{m}$) and from harmonic interaction ($\omega_{h}$), as defined by (\ref{eqn:omega_decomp}).} 
\label{tab:meanfreq}
\end{table}
 This  result explains a posteriori why a global stability analysis of the time-averaged flow  accurately predicts the frequency of the time-periodic vortex shedding in the cylinder wake \citep{B06}. \cite{ML14} recently proposed a self-consistent nonlinear model based on the marginal stability of the mean flow,  which  accurately determines  both  the mean flow and the frequency of the vortex shedding in the cylinder wake. However, the influence of higher harmonics in the frequency selection of nonlinear instability is not negligible  in all circumstances. Indeed, 
\cite{T15} recently showed  that in thermosolutal convection driven by opposite thermal and solutal gradients, oscillation frequencies of \emph{travelling} convection waves can be predicted from stability analysis of the mean flow, but not those of \emph{standing} waves. 
The application of endogeneity analysis to the thermosolutal convection problem is  not attempted here, but it  is expected to  reveal  a stronger  influence of the harmonic interactions in the frequency selection of standing waves.

\begin{figure}
\centering
\begin{tabular}{ll}
(a) & (b) \\
\includegraphics[scale=0.5]{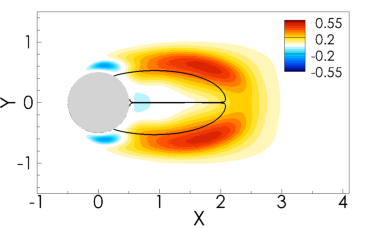} & \includegraphics[scale=0.5]{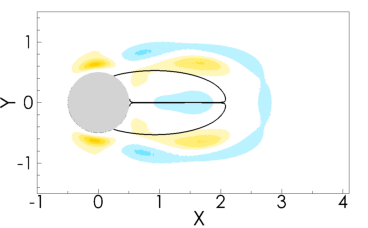} \\
(c) & (d) \\
\includegraphics[scale=0.5]{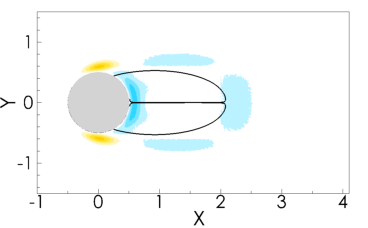} & \includegraphics[scale=0.5]{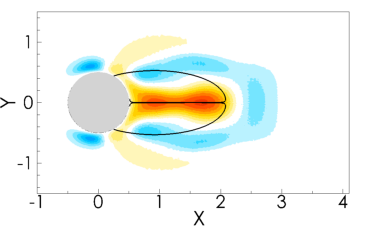}
\end{tabular}
\caption{Spatial distribution of various contributions to the endogeneity of the quasi-linear mean flow dynamics: (a) production by the mean flow, (b) convection by the mean flow, (c) diffusion and (d) pressure gradient. The black line delimits the recirculation region  in the mean flow.}  
\label{fig:endo_decompmeanlin}
\end{figure}
Finally, a decomposition of the endogeneity  component  (\ref{eqn:endo_decomp_m})    similar to the demonstration in \S \ref{sec:cylinder_linear} characterizes  the contributions of the various terms to the quasi-linear dynamics. For $\textit{Re}=80$, the contributions of production and mean flow convection to the  global  frequency are displayed in figures \ref{fig:endo_decompmeanlin}{\em a,b}, and the contributions of diffusion and pressure gradient are shown in figures \ref{fig:endo_decompmeanlin}{\em c,d}. 
The production  contribution  is dominant,  concentrated  in the shear layers of the mean recirculation region. Inside  this  region, the  action of the  pressure gradient  further increases  the frequency.  
Table \ref{tab:meanfreqlin}  summarizes all total contributions,  $\omega_{m}^{p}$, $\omega_{m}^{c}$, $\omega_{m}^{d}$, of the production, convection and diffusion  terms at various Reynolds numbers  . The  net  integral of the pressure gradient term is  always identically  zero.  At  all Reynolds numbers, the effects of  diffusion and mean flow convection approximately compensate each other, and  their balance is small  compared to the strong contribution of the production  term. 

\begin{table}
\centering
\begin{tabular}{rrccc}
$\textit{Re}$ \vline & $\omega_{m}$ \vline  & $\omega_{m}^{p} $  & $\omega_{m}^{c}$ & $\omega_{m}^{d} $   \\
\hline
$50$ \vline &  $0.8087$ \vline & $0.9997$ & $0.0023$ & $-0.1933$ \\
$75$ \vline & $0.9543$ \vline & $0.9435$ & $0.1567$ & $-0.1459$ \\
$\textbf{80}$ \vline & $\textbf{0.9768}$ \vline & $\textbf{0.9482}$ & $\textbf{0.1679}$ & $\textbf{-0.1394}$ \\
$100$ \vline & $1.0493$  \vline & $0.9999$ & $0.1696$ & $-0.1203$ \\
$125$ \vline & $1.1127$ \vline & $1.0989$ & $0.1224$ & $-0.1087$ \\
$150$ \vline & $1.1596$ \vline & $1.1721$ & $0.0853$ & $-0.0978$ \\
$175$ \vline & $1.1935$ \vline & $1.2220$ & $0.0604$ & $-0.0889$  
\end{tabular}
\caption{Decomposition of the quasi-linear contribution to the fresquency into three contributions: the production term $\omega_{m}^p$, convection term $\omega_{m}^c$ and diffusion term $\omega^d$ in the quasi-linear mean flow dynamics.} 
\label{tab:meanfreqlin}
\end{table}

\section{Conclusions}

A novel sensitivity formalism has been introduced, named the endogeneity, which allows to precisely quantify the influence of each point in the flow field on the global frequency selection and on the promotion of unstable growth, both in the context of linear temporal eigenmodes and of nonlinear global modes. Its application has been demonstrated for the Ginzburg--Landau equation and for the wake of a circular cylinder, in linear as well as nonlinear settings. The results obtained have been shown to be consistent with earlier `wavemaker' definitions, in particular the WKBJ-based saddle point criterion of \cite{CHR91} and the structural sensitivities defined by \cite{GL07}, \cite{LGP08} and \cite{H15}. The novel aspect with respect to the latter sensitivity approaches arises from the specific form of operator variations that are considered: the endogeneity characterizes the sensitivity of the eigenvalue with respect to localized operator variations that preserve the specific structure of the original operator. This sensitivity may therefore be interpreted as the \emph{local contribution} of any point in the flow field to the global eigendynamics, in terms of frequency selection and unstable growth. Contributions to these two parts of the eigenvalue are clearly distinguished, contained separately in the real and imaginary parts of the endogeneity. Further analysis of the role of individual terms of the operator follows naturally within this framework. In particular, a decomposition of a nonlinear flow operator into time-averaged and fluctuating parts gives insight into the role of quasi-linear dynamics developing on the mean flow versus the nonlinear interaction of harmonic fluctuation components. This latter part of the analysis, exemplified for the case of the cylinder wake, has important implications for the characterisation of nonlinear time-periodic flow states, in relation to recent investigations \citep{ML14, T15}.

\acknowledgments{The authors are grateful to Patrick Huerre, Jean-Marc Chomaz and Denis Sipp for their helpful comments. Lutz Lesshafft acknowledges financial support from the Agence Nationale de la Recherche under the ``Cool Jazz'' project.}

\appendix
\section{Influence of the inner product}

The question may arise whether and how the choice of an inner product different from (\ref{eqn:spatialprod}) affects the endogeneity definition. In particular, many flow problems are investigated in cylindrical coordinates, where the standard inner product includes the radial coordinate $r$ as part of the volume element. For practical purposes, the procedure is outlined here for \emph{discrete} eigenvalue problems, with a generalized inner product.

Let $\tphi_j$ be a discrete representation of the linear eigenfunction $\bphi_j$, in a discrete space where the inner product $\langle \bphi_1, \bphi_2 \rangle$ between two flow states is expressed as $\tphi_1^H \mathsfbi{Q} \tphi_2$. The matrix $\mathsfbi{Q}$ is typically diagonal, with real elements that represent volume elements of the mesh, and possibly any further weight functions. It will only be assumed here that  $\mathsfbi{Q}$ is invertible, but even this condition can be relaxed with some additional effort.

The discrete direct and adjoint eigenvalue problems are defined by
\begin{align}
-\i\omega_j \mathsfbi{B} \tphi_j &= \mathsfbi{L} \tphi_j \, ,\\
\i\omega^*_j \mathsfbi{Q}^{-1,H}\mathsfbi{B}^H \mathsfbi{Q}^{H} \tphi^\dag_j &= \mathsfbi{Q}^{-1,H}\mathsfbi{L}^H \mathsfbi{Q}^{H} \tphi^\dag_j \, . \label{eqn:discr_adj}
\end{align}
The matrix $\mathsfbi{Q}^{-1,H}$ on both sides of (\ref{eqn:discr_adj}) can be omitted. It is found that $\cphi^\dag_j=\mathsfbi{Q}^{H} \tphi^\dag_j$ is the adjoint eigenvector satisfying
\begin{equation}
  \i\omega^*_j \mathsfbi{B}^H \cphi^\dag_j = \mathsfbi{L}^H \cphi^\dag_j \, .
  \label{eqn:Hermitian_adjoint}
 \end{equation}
 
Let $\boldsymbol{\Phi}_j(\bx _k)$ be defined as the vector $\boldsymbol{\Phi}_j (\bx _k) = \delta(\bx-\bx _k)\mathsfbi{L}\tphi_j$ for convenient writing, where $\bx _k$ denotes the discrete mesh points. The endogeneity definition, written in terms of discrete vectors and operators, is then 
\begin{equation}
-\i E(\bx_k) = \langle \phi^\dagger_j, \delta(\bx-\bx_k)\,\mathcal{L}\phi_j\rangle = \tphi_j^{\dag H} \mathsfbi{Q} \boldsymbol{\Phi}_j(\bx_k) = \cphi_j^{\dag H}\boldsymbol{\Phi}_j(\bx_k) .
\label{eqn:discrete_endo}
\end{equation}

It follows from (\ref{eqn:discrete_endo}) that the endogeneity is invariant with respect to the choice of the inner product. This holds true also for the example of cylindrical coordinates: it seems unnecessary to account for the $r$ factor. Without the need to specify a $\mathsfbi{Q}$ matrix, $E(\bx_k)$ may be computed directly from the discrete adjoint eigenvector, obtained from the transpose conjugate problem (\ref{eqn:Hermitian_adjoint}). This will generally be the simplest option.

\bibliography{biblio}
\bibliographystyle{jfm}

\end{document}